\documentclass[eqsecnum,twocolumn,showpacs,preprintnumbers,amssymb,aps]{revtex4}
\usepackage{graphicx}
\usepackage{slashbox}
\usepackage{dcolumn}
\usepackage{bm}% bold math
\usepackage{latexsym,epsfig}

\begin{document}

\preprint{\today}

\title{Relativistic coupled-cluster studies of ionization potentials, lifetimes and polarizabilities in singly ionized calcium}
\vspace{0.5cm}

\author{B. K. Sahoo \protect \footnote[1]{E-mail: B.K.Sahoo@rug.nl}}
\affiliation{KVI, University of Groningen, NL-9747 AA Groningen, The Netherlands}
\author{B. P. Das}
\affiliation{Non-accelerator Particle Physics Group, Indian Institute of Astrophysics, Bangalore-560034, India}
\author{D. Mukherjee}
\affiliation{Raman Center for Atomic, Molecular and Optical Sciences,\\ Indian Association for Cultivation of Science, Kolkata-700032, India}
\date{\today}
\vskip1.0cm

\begin{abstract}
\noindent
Using the relativistic coupled-cluster method, we have calculated ionization 
potentials, E1 matrix elements and dipole polarizabilities of many low-lying
states of Ca$^+$. Contributions from the Breit interaction are given
explicitly for these properties. Polarizabilities of the ground and the first 
excited d-states are determined by evaluating
the wave functions that are perturbed to first order by the electric dipole operator
and the black-body radiation shifts are estimated from these results. We also
report the results of branching ratios and lifetimes of the first excited p-states  
using both the calculated and experimental wavelengths
and compare them with their measured values.
\end{abstract}


\maketitle

\section{Introduction}
Singly ionized calcium (Ca$^+$) is an interesting candidate in many areas 
of physics. It is especially important in astrophysics for investigating the 
radiative properties of stellar objects \cite{welty,mashonkina}. Its transition wavelengths and
electric dipole amplitudes are required to find out isotopic abundances 
\cite{hashimoto} and the energy transfers in stars \cite{welty,mashonkina}. They are also used for obtaining  
information on emission and absorption lines of the electric dipole transitions 
between the low-lying states in galaxies, interstellar gas clouds and gas 
disks surrounding the stars \cite{welty,mashonkina,persson,hobbs}. Ca$^+$ is also suitable for  
laboratory physics. Using the techniques of laser cooling and ion trapping, it has been subjected
to many precision measurements, optical frequency methodology, quantum 
processing and accurate fine structure constant measurements \cite{roos,knoop,gulde,donald,champenois,zumsteg,ito}. In these
measurements, the knowledge of polarizabilities is necessary to estimate
the black-body shift (BBS) and the Stark shift due to the external electromagnetic
fields. In our recent works, we have reported the hyperfine structure constants
and quadrupole moments in Ca$^+$ using the relativistic
coupled-cluster (RCC) method \cite{bijaya1,bijaya2,csur}. The determination of  electric dipole polarizabilities 
requires electric dipole (E1) matrix elements and excitation energies of 
all the allowed transitions. Due to the importance of these quantities,
a number of calculations based on various many-body methods including the
sum-over-states approach in the framework of the RCC theory are employed to evaluate
them \cite{patil,arora,lim,mitroy}. There are also measurements of the static 
dipole polarizability of the ground state in Ca$^+$ \cite{theodosiou,chang}, but the results do
not agree with each other. In fact, all the calculations \cite{patil,arora,lim,mitroy} differ from the recent measured value \cite{theodosiou}. Therefore, it is necessary to carry out thorough investigations 
of the role of electron correlation, higher order relativistic effects and contributions
from the two hole-two particle and the neglected one hole- one particle excited states in the
calculations of polarizabilities using an all order{\it ab initio} approach 
like the RCC theory. We have developed a novel technique
to account for the importance of 
different correlation effects in these properties for closed-shell and one-valence atomic systems
by directly obtaining the atomic wave functions perturbed to first order by the electric dipole 
operator in the framework of the RCC theory \cite{bijaya3,bijaya4}. This method avoids the 
sum-over-states approach and thereby includes 
different types of correlation effects in a rigorous manner. This theory has been employed
to determine the ground state polarizabilities in a few alkali atoms and singly 
ionized alkaline earth ions including Ca$^+$ to check the validity of the 
theory \cite{bijaya3}. Although, the theory for the tensor polarizabilities has been developed, it has not been applied to 
excited states. In this work, we calculate
the E1 matrix elements and excitation energies and employ the above approach 
to determine scalar and tensor polarizabilities of the 4S and 3D states of 
Ca$^+$. The role of the Breit interaction had not been studied in the 
earlier works which we investigate here using this {\it ab initio} method.

There have been recent measurements on the branching ratios from
the 4p $^2P_{3/2}$ state and the corresponding transition probabilities in Ca$^+$ \cite{gerritsma} which need to be theoretically
investigated. We carry out these studies using our {\it ab initio} approach and 
by combining our E1 matrix elements with the experimental wavelengths 
and compare with their corresponding experimental results. We also evaluate 
the lifetimes of the 4P-states using these results. 

The remaining part of the paper is organized as follows: In Sec. II,
we present a brief outline of the theory. This is followed in Sec. III by a discussion of the method to evaluate 
the unperturbed and the first order perturbed atomic wave functions using the 
RCC method. We then present the results and discuss the effect of correlation
on various properties in Sec. IV and in the final section we make some concluding remarks. 

\section{Theoretical approach}
The static dipole polarizability of a state $|J_0, M_0 \rangle$ is given by
\begin{eqnarray}
\alpha_0 = \alpha_0^1 + \frac {3 M_0^2 - J_0 (J_0+1)}{J_0 (2J_0-1)} \alpha_0^2
\label{eqn1}
\end{eqnarray}
where $\alpha_0^1$ and $\alpha_0^2$ are the scalar and tensor 
polarizabilities. From the angular momentum selection rule, it is obvious 
that $\alpha_0^2$ will be non-zero only for the states with $J_0 > 1/2$.
In an explicit form, the expression for the polarizability in the 
sum-over-states approach can be written as
\begin{eqnarray}
\alpha_0^i = 2 \sum_{n \ne 0} C_i \frac { |\langle J_0 || D || J_n \rangle|^2}{E_0 -E_n}, 
\label{eqn2}
\end{eqnarray}
with 
\begin{eqnarray}
C_1 &=& - \frac{1}{3(2J_0+1)}, \nonumber \\
C_2 &=& \left [ \frac{10 J_0 (2J_0-1)}{3 (J_0+1)(2J_0+1)(2J_0+3)} \right ]^2 \nonumber \\
&& \ \ \ \ \ \ \ \ (-1)^{J_0-J_n} \left \{ \matrix {J_0 & 1 & J_n \cr 1 & J_0 & 2 } \right \} \nonumber
\end{eqnarray}
and the $E$s are the energies
of the atomic states. In a single valence system, $\alpha_0^i$ can be 
divided into three parts in general as follows:
\begin{eqnarray}
\alpha_0^i = \alpha_0^i(v) + \alpha_0^i(cv) + \alpha_0^i(c),
\label{eqn3}
\end{eqnarray}
where $v$, $cv$ and $c$ inside the parenthesis represent for valence, 
core-valence and core correlation contributions, respectively. In the
sum-over-states approach, it is customary to evaluate $\alpha_0^i(v)$ by 
calculating the important valence excited states. However, contributions from 
$\alpha_0^i(cv)$ and $\alpha_0^i(c)$ are generally taken approximately in such 
an approach. On the
otherhand, it is possible to calculate $\alpha_0^i$ exactly in a particular configuration space by evaluating
the wave function that is perturbed by the electric dipole operator operator $D$ in the following manner: 

Let us rewrite Eq. (\ref{eqn2}) as
\begin{eqnarray}
\alpha_0^i &=& 2 \sum_{n \ne 0} C_i (-1)^{J_0-J_n} \frac { \langle J_0 || D || J_n \rangle \langle J_n || D || J_0 \rangle }{E_0 -E_n},
\label{eqn4}
\end{eqnarray}
which in Dirac notation can be expressed as
\begin{widetext}
\begin{eqnarray}
\alpha_0^i &=& 2 \sum_{n \ne 0} C_i (-1)^{J_0-J_n} \frac { \langle \Psi^{(0)}(J_0,\gamma) || D || \Psi^{(0)}(J_n,\gamma') \rangle \langle \Psi^{(0)}(J_n,\gamma') || D || \Psi^{(0)}(J_0,\gamma) \rangle }{E_0 -E_n} \nonumber \\
 &=& \langle \Psi^{(0)}(J_0,\gamma) || \tilde{D}_i || \Psi^{(1)}(J_0,\gamma') \rangle + \langle \Psi^{(1)}(J_0,\gamma') || \tilde{D}_i || \Psi^{(0)}(J_0,\gamma) \rangle 
\label{eqn5}
\end{eqnarray}
where $\gamma$ represents parity eigenvalue of the state $|J_0, M_0 \rangle$
and $\gamma'$ is its opposite eigenvalue and we define an effective dipole
operator as $\tilde{D}_i=C_i (-1)^{J_0-J_n} D$. Here, $|\Psi^{(1)}(J_0,\gamma')\rangle$ is the first order perturbation correction to the wave function  $|\Psi^{(0)}(J_0,\gamma)\rangle$ due to
the dipole operator $D$ and given by
\begin{eqnarray}
| \Psi^{(1)}(J_0,\gamma') \rangle &=& \sum_{n \ne 0} | \Psi^{(0)}(J_n,\gamma') \rangle \frac {\langle \Psi^{(0)}(J_n,\gamma') || D || \Psi^{(0)}(J_0,\gamma) \rangle}{E_0 -E_n}
\label{eqn6}
\end{eqnarray}
It can be equivalently written as
\begin{eqnarray}
| \Psi^{(1)}(J_0,\gamma') \rangle &=& \frac {1}{E_0 - H} \sum_{n \ne 0} | \Psi^{(0)}(J_n,\gamma') \rangle \langle \Psi^{(0)}(J_n,\gamma') || D || \Psi^{(0)}(J_0,\gamma)\rangle \nonumber \\
 &=& \frac {1}{E_0 - H} \sum_{n, \varrho=\gamma,\gamma'}| \Psi^{(0)}(J_n,\varrho) \rangle \langle \Psi^{(0)}(J_n,\varrho) || D || \Psi^{(0)}(J_0, \gamma) \rangle, \nonumber \\
\end{eqnarray}
since the matrix elements between the same parity states vanish. Applying the completeness condition, we get
\begin{eqnarray}
 (H - E_0) | \Psi^{(1)}(J_0,\gamma') \rangle &=& - D | \Psi^{(0)}(J_n,\gamma) \rangle ;
\label{eqn7}
\end{eqnarray}
\end{widetext}
the above equation can be considered as a first order perturbation equation arising from $D$. By solving the above 
equation and Eq. (\ref{eqn5}) it is possible to evaluate $\alpha_0^i$ in the 
framework of the relativistic coupled-cluster theory.

\begin{figure}[h]
\includegraphics[width=7.5cm,clip=true]{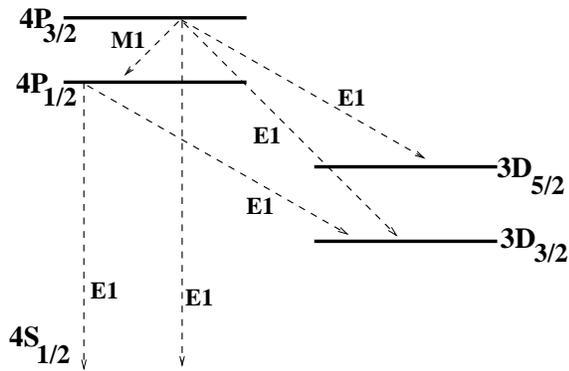}
\caption{Schematic low-lying energy level diagrams and decay channels of the P-states in Ca$^+$.}
\label{fig1}
\end{figure}
The branching ratio (BR) of a state ($f$) to a lower energy state ($i$) is defined as
\begin{eqnarray}
\Gamma_{ f \rightarrow i } = \frac {A_{ f \rightarrow i }}{\sum_i A_{ f \rightarrow i }},
\label{eqn8}
\end{eqnarray}
where $A_{ f \rightarrow i }$ is the transition probability of the corresponding transition and sum over $i$ represents total probabilities of all possible 
transitions. As shown the low-lying energy levels of Ca$^+$ in Fig. \ref{fig1}, 
the electrons from the $4p \ ^2P_{1/2}$ state will jump either to the 
$4s \ ^2S_{1/2}$ or $3d \ ^2D_{3/2}$ states due to the allowed transition with
different probabilities. Again, electrons from the $4p \ ^2P_{3/2}$ state will 
jump to the $4s \ ^2S_{1/2}$, $3d \ ^2D_{3/2}$ and $3d \ ^2D_{5/2}$ states
due to the allowed transitions and $4p \ ^2P_{1/2}$ state due to the M1 
forbidden transition. The lifetime of the state can be determined from 
\begin{eqnarray}
\tau_f = \frac {1}{\sum_i A_{ f \rightarrow i }}.
\label{eqn9}
\end{eqnarray}
 
By combining Eq. (\ref{eqn8}) and Eq. (\ref{eqn9}), it yields
\begin{eqnarray}
A_{ f \rightarrow i } = \tau_f  \Gamma_{ f \rightarrow i }.
\label{eqn09}
\end{eqnarray}
From the same or different measurements of $\tau_f$ and $\Gamma_{ f \rightarrow i }$, it is possible to estimate the corresponding $A_{ f \rightarrow i }$ for
various transitions.

The probabilities due to E1 and M1 transitions are given by
\begin{eqnarray}
A_{ f \rightarrow i }^{E1} = \frac {2.02613 \times 10^{18}}{(2J_f+1) \lambda_{ f \rightarrow i }^3} |\langle J_f || D || J_i \rangle|^2 ,
\label{eqn10}
\end{eqnarray}
and
\begin{eqnarray}
A_{ f \rightarrow i }^{M1} = \frac {2.69735 \times 10^{13}}{(2J_f+1) \lambda_{ f \rightarrow i }^3} |\langle J_f || M1 || J_i \rangle|^2 ,
\label{eqn11}
\end{eqnarray}
respectively. In the above equations, $\lambda_{ f \rightarrow i }$ is the
wavelength of the corresponding transition and it is the reciprocal of the 
excitation energy (EE).

\section{Method of calculations}
The RCC method which is equivalent to all order perturbation theory has been recently used to obtain precise results and account for 
the correlation effects in single valence systems  
\cite{bijaya1,bijaya2,csur}.
Atomic wave functions for single valence systems can be expressed in the framework of RCC 
theory as
\begin{eqnarray}
| \Psi_v^{(0)} \rangle &=& = e^T \{ 1+S_v \} | \Phi_v \rangle,
\label{eqn12}
\end{eqnarray}
where $| \Phi_v \rangle$ is the reference state constructed from the 
Dirac-Fock wave function $| \Phi_0 \rangle$ of the closed-shell configuration
$|1s^2 2s^2 2p^6 3s^2 3p^6 \rangle (\equiv |[3p^6]\rangle)$ of Ca$^+$ by 
defining $|\Phi_v \rangle= a_v^{\dagger} | \Phi_0 \rangle$ with $a_v^{\dagger}$ representing addition of a valence electron $v$. Here $T$ and $S_v$
are the RCC excitation operators which excite electrons from $|[3p^6]\rangle$
and $ a_v^{\dagger} |[3p^6] \rangle$ for the corresponding valence electron
$v$, respectively. The amplitudes of these excitation operators are solved
by 
\begin{eqnarray}
\langle \Phi^L |\{\widehat{H_Ne^T}\}|\Phi_0 \rangle = 0  \ \ \ \ \ \ \ \ \ \ \ \ \ \ \ \ \ \ \ \ \ \ \ \ \ \ \ \label{eqn13} \\
\langle \Phi_v^L|\{\widehat{H_Ne^T}\}S_v|\Phi_v\rangle = - \langle \Phi_v^L|\{\widehat{H_Ne^T}\}|\Phi_v\rangle \ \ \ \ \ \ \ \ \ \ \ \nonumber \\ \ \ \ \ \ \ \ \ \ \ \ \ \ \ \ + \langle \Phi_v^L|S_v|\Phi_v\rangle \Delta E_v , \ \ \ \ \ \
\label{eqn14}
\end{eqnarray}
with the superscript $L (=1,2)$ representing the single and double excited
states from the corresponding reference states and the wide-hat symbol over
$H_Ne^T$ represent the linked terms of normal order atomic Hamiltonian $H_N$
and RCC operator $T$. $\Delta E_v$ is the  corresponding valence electron
affinity (negative of the ionization potential (IP)) energy which is 
evaluated by
\begin{eqnarray}
 \Delta E_v = \langle \Phi_v|\{\widehat{H_N e^T}\} \{1+S_v\} |\Phi_v\rangle .
\label{eqn15}
\end{eqnarray}
The EE between two different states are determined from the difference
of their $\Delta E_v$s. In Eqs. (\ref{eqn13}) and (\ref{eqn14}) we have considered only
the single and double excitations, however we have incorporated contributions
from important triple excitations to the $\Delta E_v$ calculations. After obtaining the amplitudes for $T$, the core excitation operator, 
we solve Eqs. (\ref{eqn14}) and (\ref{eqn15}) simultaneously to obtain the
amplitudes for the $S_v$ operator. We use the 
Dirac-Coulomb-Breit Hamiltonian which is given by
\begin{eqnarray}
H  = c \vec \alpha \cdot \vec {\text{p}} + (\beta -1) c^2 + V_{nuc}(r)
 + \frac {1} {r_{12}} - \frac {\vec \alpha_1 \cdot \vec \alpha_2 } {r_{12}} + \nonumber \\
\frac {1}{2} \left \{ \frac {\vec \alpha_1 \cdot \vec \alpha_2} {r_{12}}  -
\frac {(\vec \alpha_1 \cdot \vec {\text{r}}_{12}) (\vec \alpha_2 \cdot \vec
{\text{r}}_{12})} {r_{12}^3} \right \} , \ \ \ \ \
\end{eqnarray}
where $c$ is the velocity of light, $\alpha$ and $\beta$ are the Dirac matrices
and $V_{nuc}(r)$ is the nuclear potential.

 We extend the RCC ansatz for the perturbed atomic state in the presence of the electric dipole 
operator $D$ as
\begin{eqnarray}
| \tilde{\Psi_v} \rangle &=& = e^{T+ \Omega} \{ 1+S_v + \Lambda_v \} | \Phi_v \rangle,
\label{eqn16}
\end{eqnarray}
where $\Omega$ and $\Lambda_v$ are the modified RCC operators to the 
$T$ and $S_v$ operators, respectively. Since Eq. (\ref{eqn7}) is first order in 
the $D$ operator, the above expression will reduce to
\begin{eqnarray}
| \tilde{\Psi_v} \rangle &=& = e^T \{ 1+ S_v+ \Omega ( 1+S_v) + \Lambda_v \} | \Phi_v \rangle .
\label{eqn17}
\end{eqnarray}
Now, separating the above wave function as $| \Psi_v^{(0)} \rangle$ and $| \Psi_v^{(1)} \rangle$, we get
\begin{eqnarray}
| \Psi_v^{(1)} \rangle &=& = e^T \{ \Omega ( 1+S_v) + \Lambda_v \} | \Phi_v \rangle .
\label{eqn18}
\end{eqnarray}

\begin{table}
\caption{Ionization potentials (in au) of Ca$^+$ from different works.}
\begin{ruledtabular}
\begin{center}
\begin{tabular}{lccc}
 &   &  \\
 State   &  This work & Others & Expt. \cite{sugar} \\
\hline
               &        & \\ 
4s $^2S_{1/2}$ & $-0.43627757$ & $-0.43836^a$ & $-0.43627767$ \\
               &        & $-0.43802^b$  & \\ 
               &        & $-0.436287^c$  & \\ 
3d $^2D_{3/2}$ & $-0.37396663$ & $-0.37407^a$ & $-0.37408278$ \\
               &        & $-0.37485^b$  & \\ 
               &        & $-0.373921^c$  & \\ 
3d $^2D_{5/2}$ & $-0.37361011$ & $-0.37379$ & $-0.37380626$ \\
               &        & $-0.37448^b$  & \\ 
               &        & $-0.373921^c$  & \\ 
4p $^2P_{1/2}$ & $-0.32123958$ & $-0.32217^a$ & $-0.32149667$ \\
               &        & $-0.32224^b$  & \\ 
               &        & $-0.320844^c$  & \\ 
4p $^2P_{3/2}$ & $-0.32025203$ & $-0.32111^a$ & $-0.32048108$ \\
               &        & $-0.32118^b$  & \\ 
               &        & $-0.320844^c$  & \\ 
5s $^2S_{1/2}$ & $-0.19788645$ & $-0.198293^c$ & $-0.19858760$ \\
4d $^2D_{3/2}$ & $-0.17674718$ & $-0.175144^c$ & $-0.17729894$ \\
4d $^2D_{4/2}$ & $-0.17665912$ & $-0.175144^c$ & $-0.17721142$ \\
5p $^2P_{1/2}$ & $-0.15978307$ & $-0.160060^c$ & $-0.16046888$ \\
5p $^2P_{3/2}$ & $-0.15944054$ & $-0.160060^c$ & $-0.16011231$ \\
\end{tabular}
\end{center}
\end{ruledtabular}
\label{tab1}
$^a$Relativistic MBPT(2)\cite{guet} \\
$^b$Brueckner approximation \cite{liaw} \\ 
$^c$Non-relativistic Coulomb approximation \cite{mitroy}.
\end{table}
Following Eq. (\ref{eqn7}), we solve again the amplitudes for the modified operators as
\begin{eqnarray}
\langle \Phi^L |\{\widehat{H_Ne^T} \Omega \}|\Phi_0 \rangle &=&  - \langle \Phi^L | \widehat{De^T} |\Phi_0 \rangle   \label{eqn19} \\
\langle \Phi_v^L|\{\widehat{H_Ne^T}\}\Lambda_v|\Phi_v\rangle &=& - \langle \Phi_v^L|\{\widehat{H_Ne^T} \Omega ( 1+S_v) + \widehat{De^T} \nonumber \\ && ( 1+S_v) \}|\Phi_v\rangle + \langle \Phi_v^L|\Lambda_v|\Phi_v\rangle \Delta E_v , \nonumber \\
\label{eqn20}
\end{eqnarray}
where $\widehat{De^T}$ represent again the connecting terms between $D$ and $T$
 operators. In the singles and doubles approximation, we write
\begin{eqnarray}
T &=& T_1 + T_2 \\
\Omega &=& \Omega_1 + \Omega_2 \\
S_v &=& S_{1v} + S_{2v} 
\end{eqnarray}
and
\begin{eqnarray}
\Lambda_v &=& \Lambda_{1v} + \Lambda_{2v} ,
\end{eqnarray}
where the subscripts \{1,2\} represent the single and double excitations, respectively.

Now the expression for the dipole polarizability follows as
\begin{widetext}
\begin{eqnarray}
\alpha_0^i &=& \frac {\langle \Psi_v^{(0)} | \tilde{D}_i | \Psi_v^{(1)} \rangle + \langle \Psi_v^{(1)} | \tilde{D}_i | \Psi_v^{(0) } \rangle} {<\Psi_v^{(0) }|\Psi_v^{(0) }>} \nonumber \\
&=& \frac {\langle \Phi_v |\{1+S_v^{\dagger}\} \overline{\tilde{D}_i} \{\Omega(1+S_v) + \Lambda_v \} | \Phi_v \rangle + \langle \Phi_v |\{\Lambda_v^{\dagger} + (1+S_v^{\dagger}) \Omega^{\dagger} \} \overline{\tilde{D}_i} \{1 +S_v\} | \Phi_v \rangle } {\{1+S_v^{\dagger}\} \overline{N}_0 \{1 +S_v\} }, 
\end{eqnarray}
\label{eqn21}
\end{widetext}
where we define $\overline{\tilde{D}_i}=(e^{T^{\dagger}} \tilde{D}_i e^T)$ and
$\overline{N}_0 = e^{T^{\dagger}} e^T$. Generally, both $\overline{\tilde{D}_i}$ and
$\overline{N}_0$ in the RCC approach are each represented by a non-terminating series. However, we have devised a procedure motivated by physical considerations to deal with them 
using the Wick's generalized theorem. We
evaluate first the effective zero-body, one-body, two body terms etc. systematically and 
then sandwich them (except zero-body terms) between the $S_v$, $\Lambda_v$ and their conjugate operators.
We have successfully applied this method in our earlier works \cite{bijaya1,bijaya2,csur,bijaya3,bijaya4}. The above zero-body terms, open-terms connecting
only with $\Omega$ and terms with $\Lambda_v$ give us core ($\alpha_0^i(c)$), 
core-valence ($\alpha_0^i(cv)$) and valence ($\alpha_0^i(v)$) correlation 
effects, respectively.

We also explicitly present contributions from the normalization factors
evaluating them in the following way
\begin{eqnarray}
Norm &=& \left [ \langle \Psi_v^{(0)} | \tilde{D}_i | \Psi_v^{(1)} \rangle + \langle \Psi_v^{(1)} | \tilde{D}_i | \Psi_v^{(0)} \rangle \right ]\{ \frac {1}{1+N_v} - 1 \}, \nonumber \\ &&
\label{eqn22}
\end{eqnarray}
where $N_v=\{1+S_{v}^{\dagger}\} \overline{N}_0 \{1 +S_{v}\}$.

\section{Results and Discussions}
\begin{table}
\caption{Transition matrix elements (in au) from different calculations. Recommended values from our work is given as reco.}
\begin{ruledtabular}
\begin{center}
\begin{tabular}{lcccc}
 &   &  \\
 Transition &  \multicolumn{3}{c}{This work} & Others \\
  \cline{2-4}    &   STOs & GTOs & reco &  \\
\hline
               &        & \\
4p $^2P_{1/2}\rightarrow$ 4s $^2S_{1/2}$ & 2.86 & 2.90 & 2.88 & 2.890$^a$ \\ 
               &        &  & &  2.866$^{b,l}$ \\
               &        &  & &  2.861$^{b,v}$ \\
               &        &  & &  2.898$^c$ \\
4p $^2P_{1/2}\rightarrow$ 3d $^2D_{3/2}$ & 2.50 & 2.41 & 2.40 & 2.373$^a$ \\
               &        &  & &  2.410$^{b,l}$ \\
               &        &  & &  2.244$^{b,v}$ \\
4p $^2P_{3/2}\rightarrow$ 4s $^2S_{1/2}$ & 4.02 & 4.09 & 4.03 & 4.088$^a$ \\ 
               &        &  & &  4.060$^{b,l}$ \\
               &        &  & &  4.059$^{b,v}$ \\
               &        &  & &  4.099$^c$ \\
4p $^2P_{3/2}\rightarrow$ 4p $^2P_{1/2}$ & 1.15 &1.15  & 1.15 &  \\ 
4p $^2P_{3/2}\rightarrow$ 3d $^2D_{3/2}$ & 1.12 & 1.09 & 1.09 & 1.059$^a$\\
               &        &  & &  1.076$^{b,l}$ \\
               &        &  & &  1.028$^{b,v}$ \\
4p $^2P_{3/2}\rightarrow$ 3d $^2D_{5/2}$ & 3.36 & 3.28 & 3.22 & 3.186$^a$ \\
               &        &  & &  3.234$^{b,l}$ \\
               &        &  & &  2.995$^{b,v}$ \\
               &        &  & &  3.306$^c$ \\
\end{tabular}
\end{center}
\end{ruledtabular}
\label{tab2}
$^a$Relativistic MBPT(2)\cite{guet} \\
$^{b,l}$Length gauge result with Brueckner approximation \cite{liaw} \\ 
$^{b,v}$Velocity gauge result with Brueckner approximation \cite{liaw} \\
$^c$Linearized RCC method \cite{arora}.
\end{table}

We have employed two different types of the basis functions to generate 
the atomic orbitals;
Slater type orbitals (STOs) and Gaussian type orbitals (GTOs). 
These orbitals are defined on a grid given by 
\begin{eqnarray}
r_i = r_0 \left [ e^{h(i-1)} - 1 \right ],
\label{eqn23}
\end{eqnarray}
where $i$ represents the grid points which we have taken as $750$ in total,
the step size $h$ is taken as $0.03$ in the present case and $r_0$ is
the starting point of the radial distribution from where the electron orbitals
become finite and taken as $ 2 \times 10^{-6}$. The STOs and GTOs are given by
\begin{eqnarray}
F^{STO}(r_i) = r^{n_{\kappa}} e^{-\alpha_i r_i} 
\label{eqn24}
\end{eqnarray}
and 
\begin{eqnarray}
F^{GTO}(r_i) = r^{n_{\kappa}} e^{-\alpha_i r_i^2} ,
\label{eqn25}
\end{eqnarray}
respectively. Here $n_{\kappa}$ is the radial quantum number of the orbitals 
and $\alpha_i$ is a parameter whose value is chosen to 
obtain orbitals with proper behavior inside and outside the nucleus of 
an atomic system. We further define $\alpha_i$ as
\begin{eqnarray}
\alpha_i = \alpha_0 \beta^{i-1}.
\label{eqn26}
\end{eqnarray}
We have considered $\alpha_0=0.0975$ and $\beta=1.77$ for STOs and 
$\alpha_0=0.00525$ and $\beta=2.83$ for GTOs. However, we have taken 35, 35, 30,
 30 and 25 STO and GTO basis functions to construct the s, p, d, f and g 
orbitals respectively. For RCC calculations, we have considered all
the core orbitals and virtual orbitals are considered up to 3500 au for s, p and d symmetries and 1500 au for f and g symmetries in the present calculations. In fact, it is observed that
number of virtual orbitals obtained using STOs are more in a given upper energy
limit than GTOs while bound orbital energies match well in both the cases.
To account for the contributions from the high lying orbitals in some of the 
properties that we have considered, we have estimated contributions from virtual orbitals using the 
second order many-body perturbation theory (MBPT(2)) and recommended (reco)
results are given by taking into account all these contributions.

\begin{table}
\caption{Transition probabilities (in $\times 10^6 \ s^{-1}$) in Ca$^+$.} 
\begin{ruledtabular}
\begin{center}
\begin{tabular}{lcccc}
 &   &  \\
 Transition &  \multicolumn{2}{c}{This work} & Others \\
  \cline{2-3}         & $\lambda^{cal}$   & $\lambda^{expt}$ & \\
\hline
               &        & \\
4p $^2P_{1/2}\rightarrow$ 4s $^2S_{1/2}$ & 135.240 & 134.333 & 135.26$^a$ \\
               &        &  & 132.9$^b$ \\
               &        &  & 132.5$^c$ \\
               &        &  & 136.0$^d$ \\
4p $^2P_{1/2}\rightarrow$ 3d $^2D_{3/2}$ & 9.0431 &  8.971 & 8.77$^a$ \\
               &        &  & 9.0$^b$ \\
               &        &  & 7.8$^c$ \\
               &        &  & 9.452$^d$ \\
4p $^2P_{3/2}\rightarrow$ 4s $^2S_{1/2}$ & 135.842 & 135.036 & 138.95$^a$ \\ 
               &        &  & 136.9$^b$ \\
               &        &  & 136.9$^c$ \\
               &        &  & 139.7$^d$ \\
4p $^2P_{3/2}\rightarrow$ 4p $^2P_{1/2}$ & $\sim 10^{-10}$ & $\sim 10^{-10}$ &  &  \\ 
4p $^2P_{3/2}\rightarrow$ 3d $^2D_{3/2}$ & 1.055 & 0.962 & 0.93$^a$\\
               &        &  & 0.95$^b$ \\
               &        &  & 0.87$^c$ \\
               &        &  & 0.997$^d$ \\
4p $^2P_{3/2}\rightarrow$ 3d $^2D_{5/2}$ & 8.435 & 8.419 & 8.24$^a$ \\
               &        &  & 8.5$^b$ \\
               &        &  & 7.2$^c$ \\
               &        &  & 8.877$^d$ \\
\end{tabular}
\end{center}
\end{ruledtabular}
\label{tab3}
$^a$Relativistic MBPT(2) is used  \cite{guet} \\
$^b$Length gauge result with Brueckner approximation \cite{liaw} \\ 
$^c$Velocity gauge result with Brueckner approximation \cite{liaw} \\
$^d$Linearized RCC method is employed \cite{arora}.
\end{table}
In Table \ref{tab1}, we present our IP results for the low-lying states and 
compare them with the corresponding experimental results. These results 
using STOs and GTOs were consistent. Some IPs from the excited states deviate
from the experimental results and it might be possible to improve them by increasing the 
virtual space. 
We also compare our results with other theoretical results.
Guet and Johnson had employed the relativistic MBPT(2) method to obtain their
results \cite{guet}. Liaw had employed the Brueckner approximation method to 
evaluate these energies \cite{liaw} and his results match with the above MBPT(2) results. In a recent work, 
Mitroy and Zhang have used a one electron semi-empirical core potential
in the non-relativistic framework \cite{mitroy} to estimate 
these energies which cannot distinguish the fine structure levels. Our 
method in contrast is {\it ab initio} and electron correlation effects are 
included to all orders in perturbation theory in the residual Coulomb
and Breit interaction in the one hole-one particle, two hole-two particle
and partial three hole-three particle approximation.

\begin{table*}
\caption{BRs of 4p $^2P_{1/2}$ and 4p $^2P_{3/2}$ states in Ca$^+$.}
\begin{ruledtabular}
\begin{center}
\begin{tabular}{lccccc}
 &   &  \\
 Transition &  \multicolumn{2}{c}{This work} & Others & Expt \cite{gerritsma}\\
  \cline{2-3}         & $\lambda^{cal}$   & $\lambda^{expt}$ & \\
\hline
               &        & \\
4p $^2P_{1/2}\rightarrow$ 4s $^2S_{1/2}$ & 0.9373 & 0.9374 & 0.9391$^a$ & \\
               &        &  & 0.9366$^b$ \\
               &        &  & 0.9444$^c$ \\
               &        &  & 0.9350$^d$ \\
4p $^2P_{1/2}\rightarrow$ 3d $^2D_{3/2}$ & 0.0627 &  0.0626 & 0.0609$^a$ \\
               &        &  & 0.0634$^b$ \\
               &        &  & 0.0556$^c$ \\
               &        &  & 0.0650$^d$ \\
4p $^2P_{3/2}\rightarrow$ 4s $^2S_{1/2}$ & 0.9347 & 0.9350 & 0.9381$^a$ & 0.9347(3)\\ 
               &        &  & 0.9354$^b$ & \\
               &        &  & 0.9443$^c$ \\
               &        &  & 0.9340$^d$ \\
               &        &  & 0.9357$^e$ \\
4p $^2P_{3/2}\rightarrow$ 4p $^2P_{1/2}$ & $\sim 0$ &  &  &  \\ 
4p $^2P_{3/2}\rightarrow$ 3d $^2D_{3/2}$ & 0.00726 & 0.00666 & 0.00628$^a$ & 0.00661(4) \\
               &        &  & 0.00649$^b$ \\
               &        &  & 0.00600$^c$ \\
               &        &  & 0.00667$^d$ \\
4p $^2P_{3/2}\rightarrow$ 3d $^2D_{5/2}$ & 0.0581 & 0.0583 & 0.0556$^a$ & 0.0587(2) \\
               &        &  & 0.0581$^b$ \\
               &        &  & 0.0497$^c$ \\
               &        &  & 0.0593$^d$ \\
               &        &  & 0.0643$^e$ \\
\end{tabular}
\end{center}
\end{ruledtabular}
\label{tab4}
$^a$Relativistic MBPT(2) is used  \cite{guet} \\
$^b$Length gauge result with MCDF method \cite{liaw} \\ 
$^c$Velocity gauge result with MCDF method \cite{liaw} \\
$^d$Linearized RCC method is employed \cite{arora} \\
$^e$Semi empirical \cite{mitroy} \\
\end{table*}

\begin{table}
\caption{Lifetimes (in $s$) of 4p $^2P_{1/2}$ and 4p $^2P_{3/2}$ states in Ca$^+$.}
\begin{ruledtabular}
\begin{center}
\begin{tabular}{lccccc}
 &   &  \\
 State &  \multicolumn{2}{c}{This work} & Others & Expt \\
 \cline{2-3}   &    $\lambda^{cal}$   & $\lambda^{expt}$ & \\
\hline
               &        & \\
4p $^2P_{1/2}$ & 6.931 & 6.978 & 6.94$^a$ & 7.098(20)$^d$\\
               &        &  & 7.047$^{b,l}$ & 7.07(7)$^e$ \\
               &        &  & 7.128$^{b,v}$ & 7.5(5)$^f$ \\
               &        &  & 6.875$^c$ & 6.62(35)$^g$ \\
4p $^2P_{3/2}$ & 6.881 & 6.924 & 6.75$^a$ & 6.924(19)$^d$ \\
               &        &  & 6.833$^{b,l}$ & 6.87(6)$^e$ \\
               &        &  & 6.898$^{b,v}$ & 7.4(6)$^f$ \\
               &        &  & 6.686$^c$ & 6.68(35)$^g$ \\
               &        &  &  & 6.72(2)$^h$ \\
               &        &  &  & 6.61(30)$^i$ \\
\end{tabular}
\end{center}
\end{ruledtabular}
\label{tab5}
$^a$Relativistic MBPT(2)\cite{guet} \\
$^{b,l}$Length gauge result with MCDF method \cite{liaw} \\ 
$^{b,v}$Velocity gauge result with MCDF method \cite{liaw} \\
$^c$Linearized RCC method \cite{arora}.
$^d$ Laser-beam-ion-beam technique\cite{jin} \\
$^e$ Laser-beam techniques \cite{grosselin} \\
$^f$ Beam foil technique \cite{andersen} \\
$^g$ Beam foil technique with cascade correction\cite{ansbacher} \\
$^h$ Hanle method \cite{smith} \\
$^i$ Hanle method \cite{rambow} \\
\end{table}
We present the E1 and M1 matrix elements in Table \ref{tab2}. As can be seen, results from our
STOs and GTOs differ for different transitions. We have considered contributions from virtual orbitals from both the basis functions using MBPT(2) and finally
given the consistent results as reco values. Guet and Johnson \cite{guet} have 
used B-spline basis based MBPT to obtain these results. Again, Arora {\it et al.} 
\cite{arora} have also used a B-spline basis but a linearized RCC method to obtain 
their results . Liaw \cite{liaw} has used the Brueckner approximation method to get
E1 matrix elements in both the length and velocity gauge expressions. Our 
method intrinsically contains all these many-body effects. We have also 
evaluated M1 matrix element between the 4p $^2P_{3/2}\rightarrow$ 
4p $^2P_{1/2}$ transition which is around 1.15 au; almost same with the 
3d $^2D_{5/2}\rightarrow$ 3d $^2D_{3/2}$ transition \cite{bijaya5}.

\begin{table*}
\caption{Polarizabilities (in au) of the 4s $^2S_{1/2}$, 3d $^2D_{3/2}$ and 3d $^2D_{5/2}$ states in Ca$^+$.}
\begin{ruledtabular}
\begin{center}
\begin{tabular}{lccccccc}
 &   &  \\
 State &  \multicolumn{4}{c}{This work} & \multicolumn{2}{c}{Others} & Expt\\
 \cline{2-5} \cline{6-7}  &   \multicolumn{2}{c}{GTOs (reco) } & \multicolumn{2}{c}{STOs} &  & & \\
       & $\alpha_0^1$ & $\alpha_0^2$ & $\alpha_0^1$ & $\alpha_0^2$ & $\alpha_0^1$ & $\alpha_0^2$ & $\alpha_0^1$ \\
\hline
               &        & \\
 4s $^2S_{1/2}$ & 73.002 &  & 74.342 & & 76.1(1.1)$^a$ & & 70.89(15)$^e$ \\ 
               &        &  & &  & 75.49$^b$ & & 75.3(4)$^f$ \\
               &        &  & &  & 70.872$^c$ & & 72.5(19)$^g$ \\
               &        &  & &  & 70.6$^d$ \\
 3d $^2D_{3/2}$ & 28.504 & $-15.870$ & 31.604 & $-17.678$ & 32.73$^c$ & $-25.20$$^c$ & \\
               &        &  & &  & 25.4$^d$ \\
 3d $^2D_{5/2}$ & 29.307 & $-22.492$ & 32.531 & $-25.516$ & 32.0(1.1)$^a$ & $-24.5(4)^a$ &  \\
               &        &  & & & 32.73$^c$ & $-25.20$$^c$ &  \\
               &        &  & & & 25.4$^d$ &   &  \\
\end{tabular}
\end{center}
\end{ruledtabular}
\label{tab6}
$^a$Linearized RCC method is employed \cite{arora} \\
$^b$Non-relativistic sum-over-oscillator strengths approach \cite{mitroy} \\
$^c$Non-relativistic sum-over-oscillator strengths approach \cite{theodosiou} \\
$^d$Non-relativistic sum-over-oscillator strengths approach \cite{barklem} \\
$^e$Lifetime measurements and oscillator strengths of \cite{theodosiou} \\
$^f$ \cite{chang} \\
$^g$ \cite{chang}.
\end{table*}
Using the above matrix elements, we determine the transition probabilities and
present them in Table \ref{tab3}. We have followed two approaches to determine
them. First we have considered energies from our calculations and derived
wavelengths ($\lambda^{cal}$) to obtain the {\it ab initio} results. In the other case, we use
our matrix elements with the experimental wavelengths ($\lambda^{expt}$). 
Although the M1 transition amplitude from the 4p $^2P_{3/2}$ state is finite
due to a very small fine structure splitting, the corresponding transition probability is almost negligible. We have also
compared our results with other {\it ab initio} and semi-empirical results
in the same table. In a recent work, Gerritsma {\it et al.} \cite{gerritsma}
have measured BRs (we discuss these results below in detail) from the
4p $^2P_{3/2}$ state and obtain various transition probabilities from this
state by combining their results with the lifetime measurements
as given by Eq. (\ref{eqn09}). Our results with $\lambda^{expt}$ match well with their results.

\begin{table*}
\caption{Contributions from DF and important RCC terms for the dipole polarizabilities. $\overline{\tilde{D}_i}$ and $\overline{\tilde{D}_i} \Omega$ are the core ($\alpha_0^i(c)$) and core-valence ($\alpha_0^i(cv)$) correlation effects, respectively. The remaining terms except $Norm$ represent the valence correlation contributions. $Norm$ gives the correction due to the normalization of the wave functions.}
\begin{ruledtabular}
\begin{center}
\begin{tabular}{lccccc}
Terms  & 4s $^2S_{1/2}$  & \multicolumn{2}{c}{3d $^2D_{3/2}$}  & \multicolumn{2}{c}{3d $^2D_{5/2}$} \\
\hline \\
DF & 96.201  & 91.487 & $-59.261$ & 89.340 & $-81.330$ \\
 & & \\
$\overline{\tilde{D}_i}$ & 2.730  & 2.730 & $-0.178$ & 2.730 & $-0.178$ \\
$\overline{\tilde{D}_i} \Omega + cc$ & 0.038 & 0.151 & $-0.245$ & 0.268  & $-0.268$ \\
$\overline{\tilde{D}_i} \Lambda_{1v} + cc$ & 77.283 & 32.427 & $-18.317$ & 33.162 & $-27.573$ \\
$\overline{\tilde{D}_i} \Lambda_{2v} + cc$ & $-1.865$ & $-0.927$ & $-0.271$ & $-0.910$ & 0.618 \\
$S_{1v} \overline{\tilde{D}_i} \Lambda_{1v} + cc$ & $-2.543$  & $-5.097$ & 2.876 & $-5.161$ & 4.277 \\
$S_{2v} \overline{\tilde{D}_i} \Lambda_{2v} + cc$ & $-2.017$ & $-0.298$ & 0.105 & $-0.289$ & 0.132 \\
Others & 0.130 & 0.161 & $-0.226$ & 0.161 & $-0.050$ \\
$Norm$ & $-0.754$ & $-0.643$ & 0.386 & $-0.654$ & 0.550 \\
\end{tabular}
\end{center}
\end{ruledtabular}
\label{tab7}
Note: Subscripts $1v$ and $2v$ represent the valence contributions due to the singly and doubly excited states, respectively.
\end{table*}
Using the above transition probabilities, we determine BRs from different calculations 
and present them in Table \ref{tab4}. These results are compared with the
recently measured values of the 4p $^2P_{3/2}$ state \cite{gerritsma}.
As presented in this table, our results with $\lambda^{expt}$ match
well the measurements. When we evaluate BRs for the 4p $^2P_{1/2}$ and 4p $^2P_{3/2}$ states due to 3d states using the relation
\begin{eqnarray}
\Gamma_{ f \rightarrow i } = \frac {A_{ f \rightarrow i }}{\sum_{i=\text{3d} \ ^
2D_{3/2}, \text{3d} \ ^2D_{5/2}}  A_{ f \rightarrow i }},
\label{eqn80}
\end{eqnarray}
it gives as 14.97 and 14.40, respectively, which are not within the error bar of the existing experimental result \cite{gallagher} and hence require further measurements for verification.

\begin{table}
\caption{Breit interaction contributions to various properties.}
\begin{ruledtabular}
\begin{center}
\begin{tabular}{lcc}
 &    \\
 State & Results \\
\hline
 &    \\
IP (au)    & \\
 4s $^2S_{1/2}$ & 0.00003056 \\
 3d $^2D_{3/2}$ & $-0.00034491$ \\
 3d $^2D_{5/2}$ & $-0.00027547$ \\
 4p $^2P_{1/2}$ & 0.00005439 \\
 4p $^2P_{3/2}$ & 0.00002354 \\
 5s $^2S_{1/2}$ & 0.00001018 \\ 
 4d $^2D_{3/2}$ & 0.00003778 \\
 4d $^2D_{5/2}$ & 0.00001114 \\ 
 5p $^2P_{1/2}$ & 0.00001729 \\ 
 5p $^2P_{3/2}$ & 0.00000674 \\
 &    \\
E1 elements (au)    & \\
4s $^2P_{1/2} \rightarrow $ 4s $^2S_{1/2}$ & 0.001 \\
4s $^2P_{1/2} \rightarrow $ 3d $^2D_{3/2}$ & $-0.012$ \\
4s $^2P_{3/2} \rightarrow $ 4s $^2S_{1/2}$ & 0.001 \\
4s $^2P_{3/2} \rightarrow $ 3d $^2D_{3/2}$ & $-0.002$ \\
4s $^2P_{3/2} \rightarrow $ 3d $^2D_{5/2}$ & $-0.005$ \\
 &    \\
Polarizability (au)    & $\alpha_0^1$ & $\alpha_0^2$ \\
 4s $^2S_{1/2}$ & $-0.011$ \\
 3d $^2D_{3/2}$ & $-0.384$ & 0.226 \\
 3d $^2D_{5/2}$ & $-0.499$ & 0.415 \\
\end{tabular}
\end{center}
\end{ruledtabular}
\label{tab8}
\end{table}
There are a number of experimental lifetime measurements available for the
4p $^2P_{1/2}$ and 4p $^2P_{3/2}$ states \cite{jin,grosselin,andersen,ansbacher,smith,rambow} using beam laser, beam foil, beam foil with cascade correction  
and Hanle techniques. Among them the laser-beam-ion-beam spectroscopy
by Jin and Church \cite{jin} results are the most precise.
Substituting our transition probabilities in Eq. (\ref{eqn9}), we
obtain the lifetimes of the 4p $^2P_{1/2}$ and 4p $^2P_{3/2}$ states 
as 6.931s and 6.881s with $\lambda^{cal}$, respectively, where as
6.979s and 6.924s with $\lambda^{expt}$, respectively. Other calculations
based on the above discussed results also predict results close to ours. In fact,
our result 6.924s of lifetime of the 4p $^2P_{3/2}$ state is in good agreement 
with the experimental results. 

Using the same wave functions used to obtain the above properties and 
solving Eq. (\ref{eqn7}), we obtain the static dipole polarizabilities of 
the 4s $^2S_{1/2}$, 3d $^2D_{3/2}$ and 3d $^2D_{5/2}$ states with STOs and
GTOs and they
are presented in Table \ref{tab6}. The dipole polarizabilities 
for the ground state from STOs and GTOs are in good agreement, but
the 3d state dipole polarizabilities differ by 4\%. Since we were able to
generate less number of virtuals using GTOs in a given energy upper
bound, the convergence of these results were checked with virtual 
orbitals with higher energies which was not possible for STOs due to
the computational limitation. Therefore, we consider our results from
GTOs are more accurate than results from STOs. There are also a number of theoretical
calculations available on both the ground and 3d excited states including
our previous work and references therein \cite{bijaya3,mitroy,arora,patil,theodosiou,barklem}.  
We had just carried out the ground state polarizability calculation in Ca$^+$ along
with other atomic systems in the earlier work \cite{bijaya3} to verify the 
validity of the method that was proposed for the first time. In the present
case, we have investigated the accuracy of the wave functions in Ca$^+$ to 
obtain IPs and E1 matrix elements which are the ingredients to evaluate 
accurate dipole polarizabilities. In fact, the correlation behavior in the
3d state dipole polarizabilities is not discussed in the literature.
Patil and Tang \cite{patil} had used multipolar-matrix elements based in 
the non-relativistic approximation to obtain the 4s $^2S_{1/2}$ state dipole 
polarizability. This has got both the summation and integration approach
over the intermediate states from different orbital quantum numbers. Using
Coulomb approximation with the Hartree-Slater core calculations, Theodosiou
{\it et al}. \cite{theodosiou} had reported the dipole polarizability of the 
same state. Their result differs from ours and it seems as  though they have not taken 
core-correlation into account. Recently, Arora {\it et al}. \cite{arora} and 
Mitroy and Zhang \cite{mitroy} have also evaluated dipole polarizabilities 
based on the sum-over E1 matrix elements and oscillator strengths between
different states. The main differences in their results and ours is that 
they have estimated core (neglected for tensor polarizability) and core-valence correlation effects approximately
whereas we have used the first order perturbed RCC method to evaluate them. Contributions
from the continuum and doubly excited states with configurations like
$[4p^5]nsms$ ($n \ne m$, with $n,m$ being principal quantum numbers) which
are also important for the dipole polarizability calculations of the 
states have been considered by us. They are implicitly accounted for in the present work by evaluating 
the first order perturbed wave functions due to the electric dipole operator. We have also corrected
our results due to the normalization of the wave functions.
In Table \ref{tab7}, we present contributions from the DF and the individual RCC terms
obtained using GTOs. The differences between these two results give the 
correlation contributions associated in evaluating these quantities.
It is evident from our studies that correlation effects
in the 3d-states are more than 50\% while it is about 20\% in the 
4s $^2S_{1/2}$ state. The $\alpha_0^i(c)$ and $\alpha_0^i(cv)$ contributions 
are found to be smaller for the scalar dipole polarizability than the 
previously estimated results. We also present these contributions for the
tensor polarizabilities which were neglected earlier. Contributions due to
the doubly excited states and normalization corrections cannot be neglected
in precision calculations. There are three experimental results for the 
ground state dipole polarizability \cite{theodosiou,chang} , but they do 
do not match with each other. Although the result given by Theodosiou {\it et al}. \cite{theodosiou} 
is the latest, but our results are close to Chang \cite{chang}.

The frequency shift (in Hz) due to a black-body (BBS) due to the frequency-dependent 
electric field at temperature $T=300K$ by neglecting the dynamic correction in 
the 4s $^2S_{1/2}\rightarrow$ 3d $^2D_{5/2}$ transition is approximated by
\cite{itano0}
\begin{eqnarray}
\Delta \nu = \frac{1}{2} (831.9 \text{V}/\text{m})^2 \left ( \frac{T(K)}{300} \right )^4 [\alpha_0^1(4s) - \alpha_0^1( 3d_{5/2}) ] .
\end{eqnarray}
By substituting our results in the above expression, we obtain $\Delta \nu=0.376$Hz which is in agreement with 0.38(1)Hz by Arora {\it et al}. \cite{arora} and 
0.368Hz by Mitroy and Zhang \cite{mitroy}. This also supports the measured
value 0.39(27)Hz \cite{champenois}. The agreement between different 
calculations is mainly due to the cancellation of the results of 4s $^2S_{1/2}$ 
and 3d $^2D_{5/2}$ states. 

In Table \ref{tab8}, we present the contributions from the Breit interaction 
to different properties. These contributions are smaller in these properties
than in the hyperfine structure constants which were reported recently 
\cite{bijaya2}. In contrast to the hyperfine constants where the Breit interaction
contributes more to the 4s $^2S_{1/2}$ state, it is larger in the 3d states 
than the ground state in the these properties.

\section{Conclusion}
We have employed the relativistic coupled-cluster method with two
different basis functions to study ionization potentials, electric dipole
matrix elements and dipole polarizabilities in the singly ionized calcium.
We have also evaluated transition probabilities, branching ratios and
lifetimes of the first excited p-states using these results.
By determining the first order perturbed wave function due to the electric
dipole operator, we obtain {\it ab initio} results for the static dipole 
polarizabilities in the ground and first excited d-states. Black-body shift
in the 4s $^2S_{1/2}\rightarrow$ 3d $^2D_{5/2}$ transition has been evaluated
using these results and compared with the other available results. Contributions
from the Breit interaction to the above properties have been studied for the
first time in singly ionized calcium.

\section{Acknowledgment}
This work is supported by NWO under VENI fellowship grant with project number
680-47-128. DM thanks DST (New Delhi) for the award of the J. C. Bose 
Fellowship and Jahawarlal Center for Advanced Scientific Research, Bangalore 
for conferring him honorary professorship. We thank C. Roos for many useful discussions. The computations 
were carried out using the Tera-flop Super computer, Param Padma in C-DAC, Bangalore.

\end{document}